\documentclass{article}
 
\usepackage[utf8]{inputenc}
\usepackage{amsmath}
\usepackage{amssymb}
\usepackage{amsfonts}
\usepackage{float}

\usepackage{graphicx} \usepackage{subcaption}
\usepackage{tabularx}
\usepackage{booktabs} 

\usepackage{algorithm}
\usepackage{algpseudocode} 
\usepackage{hyperref}
\usepackage[capitalise]{cleveref} 

\usepackage{todonotes} 
\usepackage{siunitx}

\usepackage[affil-it]{authblk}  

\usepackage{enumitem} 

\usepackage[displaymath]{lineno}

\newcommand{\change}[1]{#1}

\makeatletter
\let\LN@align\align
\let\LN@endalign\endalign
\renewcommand{\align}{\linenomath\LN@align}
\renewcommand{\endalign}{\LN@endalign\endlinenomath}
\let\LN@alignat\alignat
\let\LN@endalignat\endalignat
\renewcommand{\alignat}{\linenomath\LN@alignat}
\renewcommand{\endalignat}{\LN@endalignat\endlinenomath}
\let\LN@gather\gather
\let\LN@endgather\endgather
\renewcommand{\gather}{\linenomath\LN@gather}
\renewcommand{\endgather}{\LN@endgather\endlinenomath}
\makeatother

\newcommand{\ulp}{\text{ulp} } 
\newcommand{\Sp}{S_{+}}
\newcommand{\Sm}{S_{\text{-}}}
\newcommand{\Spm}{S_{\pm}}
\newcommand{\N}{\mathbb{N}}
\newcommand{\R}{\mathbb{R}}
\newcommand{\norm}[1]{\left\lVert#1\right\rVert}

\author[1]{Raju, S.}
\author[1]{Gr\"unding, D.}
\author[1]{Mari{\'c}, T.}
\author[1]{Bothe, D.\thanks{Corresponding author. Email: \href{mailto:bothe@mma.tu-darmstadt.de}{bothe@mma.tu-darmstadt.de}}}
\author[1]{Fricke, M.}
\affil[1]{Institute for Mathematical Modeling and Analysis, TU Darmstadt, Germany}
\date{}                     

\title{Computing hydrodynamic eigenmodes of channel flow with slip - A highly accurate algorithm}

\begin{document}

\maketitle

\begin{abstract}
The transient start-up flow solution with slip is a useful tool to verify Computational Fluid Dynamics (CFD) simulations. However, a highly accurate, open-source black box solution does not seem to be available. Our method provides a fast, automated, and rigorously verified open-source implementation that can compute \change{the hydrodynamic eigenmodes of a two-dimensional channel flow} beyond the standard floating-point precision. \change{This allows for a very accurate computation of the corresponding Fourier series solution.} We prove that all roots are found in all special cases for the general flow problem with different slip lengths on the channel walls. \change{The numerical results confirm analytically derived asymptotic power laws for the leading hydrodynamic eigenmode and the characteristic timescale in the limiting cases of small and large slip. }\\
\\
The code repository including test cases is publicly available here\\
  \url{https://git.rwth-aachen.de/fricke/start-up-flow}
\end{abstract}

 \section{Introduction} \label{sec:introduction}
\change{The no-slip condition for the flow of a viscous liquid along a solid boundary is one of the corner stones of classical hydrodynamics. However, it is an assumption that cannot be derived from first principles \cite{Lauga2007} and there are indeed examples of flow configurations where it seems to break down. Maybe the most prominent example is the ``moving contact line paradox'' in dynamic wetting \cite{Huh1971}. Here one considers the two-phase flow of a viscous fluid displacing a second fluid to form a dynamic ``contact line'' moving along the solid boundary. It has been shown, that this fundamental process, which is present in typical dynamic wetting flows such as drop spreading or capillary rise, can not be described in the classical hydrodynamic model based on the no-slip condition \cite{Huh1971}. In particular, a singularity emerges in the viscous dissipation which would require an infinite force to move the contact line. There have been numerous approaches to resolve this apparent contradiction to the experimental observation of dynamic wetting flows (see, e.g., \cite{Gennes1985,Shikhmurzaev2008,Bonn2009}). One important example is the introduction of a tangential slip boundary condition at the solid boundary. Originating from the classical work by Navier in the 19th century, the tangential slip at the solid boundary is modeled proportional to the viscous stress, i.e.}
\begin{align}
\lambda u + \eta \, \partial_y u = 0 \quad \Leftrightarrow \quad u + L \, \partial_y u = 0,
\end{align}
\change{where $u$ is the velocity component along the solid boundary and $y$ is the coordinate normal to the solid boundary. The ratio of the dynamic viscosity $\eta$ and the friction coefficient $\lambda$ has the dimension of length and is called the ``slip length'' $L=\eta/\lambda$. Note that this boundary condition can also be motivated as a closure relation for the entropy production along the solid boundary \cite{Ren2007}. In particular, the second law of thermodynamics only allows non-negative values of the slip length. It has been shown that any $L>0$ allows for a moving contact line (with integrable viscous dissipation) \cite{Huh1977} and thereby resolves the moving contact line paradox\footnote{It should be noted that a weak singularity at the moving contact line remains if the slip length is finite; see \cite{Fricke2019,Huh1977} for more details.}. Hence, even a tiny amount of slip can change the macroscopic prediction of the model drastically. This is why the slip parameter is of such a crucial importance in dynamic wetting.}\\
\\
\change{Remarkably, there is experimental evidence that tangential slip is present and measurable even in single-phase flows where its impact on macroscopic quantities is, however, typically much smaller; see \cite{Neto2005,Lauga2007,VegaSanchez2021} for a comprehensive review of the topic. For example, the volume flow rate of a Newtonian liquid through a cylindrical channel of radius $R$ and length $l$ is proportional to \cite{Neto2005}}
\begin{align}
Q \propto \frac{\Delta p \, R^4}{\eta l} \left(1 +  \frac{4 L}{R} \right). 
\end{align}
\change{Hence, the correction due to slip is very small if the slip length is much smaller than the channel radius. In this case, it is hard to distinguish the slip effect from the variation that comes from the uncertainties in the other physical parameters such as the radius. Remarkably, the method introduced recently by Chen et.\ al \cite{Chen2015} can unambiguously distinguish boundary slip from other effects by analyzing the spectrum of hydrodynamic eigenmodes in molecular dynamics simulations of a channel flow.  At the same time, the method allows to independently determine the position of the hydrodynamic boundary. This shows that the spectrum of eigenmodes is a rich source of information about the physical system under consideration.}\\
\\
\change{Careful experiments show that the slip length (if it is measurable) lies typically in the range of a few nanometers; see \cite{VegaSanchez2021} for an overview. There are some exceptions like lubricated (or lubricant-infused) surfaces where the effective slip length (measured at the interface between the working fluid and the lubricant) can be much larger than the thickness of the lubricant layer; see, e.g, \cite{VegaSanchez2022}. In this case, the effective slip can significantly reduce the amount of energy required to transport the liquid through the channel. In the case of dynamic wetting, the very small length scale of slip introduces substantial challenges regarding computational feasibility of numerical simulations. The introduction of a nanoscale parameter (in this case the slip length) requires to bridge many orders of magnitudes up to the process scale which can be millimeters or even larger.}\\
\\
On the macroscopic scale, the slip boundary condition \change{can be applied as an effective boundary condition} to incorporate the influence of roughness or porousity \cite{Wang2003a}, the interaction of dense particulate emulsions (suspensions, foams, polymer solutions) \cite{Yoshimura1988}, interaction with chemically treated hydrophobic surfaces \cite{Tretheway2002}, or the interaction with rarefied gases \cite{Sharipov1998,Avramenko2015}. On intermediate scales in the micro to nanometer regime, the slip boundary condition is used in the layout of micro-electro-mechanical systems \cite{GadelHak2004,Castelloes2007}. The applications even reach to atomic length scales to account for wall friction effects \cite{Majumder2005}.\\
\\
\change{The instationary flow of a Newtonian liquid in a two-dimensional channel in the presence of boundary slip has been studied based on Fourier series expansions by Matthews and Hastie \cite{Matthews2012}. The Fourier series solution can be used, e.g., as a verification for an implementation of the Navier slip boundary condition in a CFD method. However, in practice this requires the accurate computation of the hydrodynamic eigenmodes and the corresponding Fourier coefficients describing the initial condition.  The eigenmodes (or eigenvalues) and expansion coefficients are frequently computed for selected cases ``by hand'' and the accuracy of the results is often omitted. Therefore, an automated, freely available and well-tested implementation seems to be required. Such a black-box implementation can then be used to produce reliable reference data for continuum and Lattice-Boltzmann simulations \cite{Avramenko2015,Gruending2019, John2002,Matthews2012}. Moreover, we show below that gives interesting insights into the spectrum of hydrodynamic eigenmodes and characteristic time-scales of instationary channel flow.}\\
\\
\change{The remainder of this article is organized as follows:} In \cref{sec:analyticSolution} the series solution for a starting channel flow with slip boundary conditions from \cite{Matthews2012} is recalled and different limit cases are outlined. \change{In particular, asymptotic formulas are derived in the limit of very small and very large slip.} The implemented algorithm including special cases is described in \cref{sec:algorithm}. \change{We provide both a Python and a C++-implementation of the algorithm in an open research data repository\footnote{\url{https://git.rwth-aachen.de/fricke/start-up-flow}}}. The two implementations are verified against each other and highly resolved CFD data in \cref{sec:results}. \change{Finally, we study the characteristic time-scale for the start-up flow based on the previously derived asymptotic scaling laws and compare with the time-scale computed from the Fourier series expansion.}

\section{Mathematical model}
\label{sec:analyticSolution}
\subsection{Channel flow with non-equal wall slip}

In the following, we consider the starting flow of an incompressible Newtonian liquid with constant temperature which is moving between two infinitely extended parallel plates. The origin of the coordinate system is placed in the center of the channel where the $x$-coordinate points in the direction of the flow and the $y$ coordinate is orthogonal to the channel walls.
Note that a different choice of coordinate systems leads to a different form of the solution and a slightly different characteristic equation to be solved for the series coefficients.
The overall algorithm developed below can also be adapted for this difference, though in the following, we will consider the case with a coordinate system in the center of the channel. We consider a Stokes flow regime leading to
\begin{align} \label{eq:momEq}
 \rho \partial_t u  &= \mu \partial_{yy} u + G
\end{align}
for the momentum conservation equation. The velocity $u$ is pointing in the channel direction, $\rho$ is the liquid density, $\mu$ is the dynamic viscosity of the liquid and $G= -\nabla p$ is the constant pressure gradient in $x$ direction. As initial conditions
\begin{alignat}{2}\label{eq:initial_cond}
 u&=0 \qquad &&\text{at} \quad t=0 ~\text{for} ~ -R \leq y \leq R
\end{alignat}
are used. At both wall boundaries, a Navier slip boundary condition 
\begin{alignat}{2}\label{eq:navier_slip_bc}
 \mp L^\pm \partial_y u &= u \quad &&\text{at} \quad y=\pm R ~\text{for}~ t \geq 0
\end{alignat}
is applied. \change{To ensure consistency with the second law of thermodynamics, we require $L^\pm \geq 0$. Note that the no-slip condition is recovered in the limit $L^\pm \rightarrow 0$. Conversely, the limit $L^\pm \rightarrow \infty$, i.e.}
\[ \partial_y u = 0 \quad \text{at} \ y = \pm +1 \]
\change{is known as the ``free-slip'' condition.}\\
\\
To reduce the number of parameters and represent the results in the following sections in a compact format, we scale \cref{eq:momEq} as well as the initial and boundary conditions \eqref{eq:initial_cond}-\eqref{eq:navier_slip_bc} by introducing the following dimensionless variables
\begin{align}\label{eq:scales}
  u^* = u /\frac{GR^2}{2\mu}, \qquad y^* = y/R, \qquad t^*= t/\frac{\rho R^2}{\mu}, \qquad \Spm = L^\pm/R, \end{align}
Here, $u^*$ is the velocity, $y^*$ is the cross-section coordinate, $t^*$ is time and the slip lengths are described by the parameters $\Spm$, which all are dimensionless.
This yields the dimensionless form of \cref{eq:momEq}
\begin{align}\label{eq:pde} 
 \partial_t^* u^* = \partial_{y^*y^*} u^* + 2
\end{align}
with initial and boundary conditions
\begin{alignat}{2}
 u^*&=0  \qquad &&\text{at} \quad t^*=0 ~\text{for} ~ -1 \leq y^* \leq 1, \label{eq:init}\\
 \mp \Spm \partial_{y^*} u^* &= u^*  &&\text{at} \quad y^*=\pm 1 ~\text{for}~ t^* \geq 0. \label{eq:bcs}
\end{alignat}
From here on we drop the $*$-notation and assume that we only deal with dimensionless quantities if not stated otherwise. A series solution of \cref{eq:pde}, \cref{eq:init}, and \cref{eq:bcs} is, e.g., available in \cite{Matthews2012}. \change{We shall briefly recall the mathematical derivation in the following.}

\paragraph{Decomposition of the solution:} 
\change{The full solution is split into a stationary and transient contribution $\overline{u}$ and $\tilde{u}$, respectively, in the form}
\begin{align} \label{eq:solNavierSlip}
 u(t, y) = \overline{u}(y) - \tilde{u}(t,y),
\end{align}
\change{where }
\begin{align} \label{eq:usNavierSlip}
\overline{u}(y) = 1-y^2 + \frac{2(\Sp + \Sm) + 4\Sp \Sm}{\Sp + \Sm + 2} - \frac{2(\Sm - \Sp)}{\Sp + \Sm +2} y
\end{align}
\change{solves the stationary problem}
\begin{align*}
0 = \partial_y^2 \bar{u} +2
\end{align*}
\change{subject to the boundary conditions \eqref{eq:navier_slip_bc}. Consequently, the transient part $\tilde{u}$ satisfies}
\begin{equation}\label{eq:ibvp-transient-part}
\begin{aligned}
\partial_t \tilde{u} = \partial_y^2 \tilde{u}, \quad t>0, \ y \in (-1,1)\\
\tilde{u}(0,y) = \bar{u}(y), \quad y \in [-1,1],\\
\tilde{u} \pm S^\pm \partial_y \tilde{u} = 0, \quad t>0, \ y \in (-1,1).
\end{aligned}
\end{equation}
\change{The transient part can be obtained using separation of variables based on the ansatz}
\[ \tilde{u}(t,y) = T(t) Y(y) \]
\change{leading to the eigenvalue problem}
\begin{align*}
T'(t) + \lambda T = 0,\\
Y''(y) + \lambda Y = 0, \quad Y(\pm1) \pm S^\pm Y'(\pm1) = 0
\end{align*}
\change{for the linear operators $\partial/\partial t$ and $-\partial^2/\partial y^2$. Restricting our attention to bounded solutions, we may require non-negative eigenvalues}
\[ \lambda_n = k_n^2 \quad \text{for} \ k_n \geq 0, \]
\change{i.e., we define $k_n$ as the  square root of the $n$-th eigenvalue $-\partial^2 / \partial y^2$ subject to the Navier slip boundary condition. Note that the decay time $\tau_n$ of an eigenmode with eigenvalue $\lambda_n = k_n^2$ is proportional to $1/k_n^2$:}
\[ e^{-k_n^2 \tau_n} = \alpha  \]
\change{Choosing (by convention) $\alpha=0.1$, we obtain}
\begin{align} 
\tau_n = \frac{\ln 10}{k_n^2} 
\end{align}
\change{for the non-dimensional time that the $n$-th eigenmode takes to decay to $10\%$ of the initial value. Consequently, the smallest positive eigenvalue $\lambda_1 = k_1^2$ corresponds to the largest decay time.}

\paragraph{Characteristic equation and Fourier series expansion:}
\change{Evaluation of the Navier slip condition at $y=-1$ leads to eigenfunctions of the form}
\[ Y(y) = \sin(k_n(y+1)) + S^-k_n \cos(k_n(y+1) \]
\change{and the characteristic equation}
\begin{align}\label{eq:charEq}
 (1-\Sp \Sm k^2) \sin(2k) + k(\Sp + \Sm) \cos(2 k) = 0
\end{align}
\change{which results from the Navier slip condition at $y=1$. Any non-negative solution $k_n$ of \eqref{eq:charEq} yields an eigenfunction (called a ``hydrodynamic eigenmode'') with eigenvalue $\lambda_n = \sqrt{k_n}$.}\\
\\
The full solution to the initial-boundary value problem \eqref{eq:ibvp-transient-part} can then be obtained from a Fourier series expansion
\begin{align} \label{eq:utNavierSlip}
 \tilde{u}(t,y) = \sum_{n=1}^\infty  A_n \left(\sin(k_n(y+1)) + S^-k_n \cos(k_n(y+1))\exp(-k_n^2 t) \right) .
\end{align}
The formula for the coefficient $A_n$ from Matthews and Hastie \cite{Matthews2012} can be significantly simplified by multiple insertions of the characteristic equation \cref{eq:charEq}, yielding
\begin{align}\label{eq:An}
 A_n 
 = 
\frac{8\sin(k_n) \left(\sin(k_n) + \Sm k_n \cos(k_n)\right) ( k_n^2\Sp^2+1)}
{k_n^3\left(2\Sp^2\Sm^2k_n^4 + (\Sp^2 (\Sm + 2) + \Sm^2(\Sp + 2) ) k_n^2 + \Sp + \Sm +2\right)}.
\end{align}
Note that $A_n$ explicitly depends on $\Sp$, $\Sm$, and $k_n$. In contrast, no such explicit formula for the coefficients $k_n$ is available in the general case. \change{Except for some limiting cases (see below), these coefficients have to be obtained by some sort of numerical algorithm, which is the main focus of the present article.}\\
\\
In \cite{Matthews2012} the characteristic equation is \change{used} in the form 
\begin{align}\label{eq:charFunctionMatthews}
 \tan(2k) - \frac{k (S^+ + S^-)}{k^{2} S^+ S^- -1} = 0. \end{align}
However, as shown below, this form of the equation is not applicable in the general case. Note that the solutions of \cref{eq:charEq} are also required for the analytic start-up solution in a Couette flow setup \cite{Ng2017}. We also give a factorization of the  characteristic equation \cref{eq:charEq} to be used below:
\begin{align}\label{eq:charEqFactored}
 (\cos k - \Sp k\sin k)&(\sin k + \Sm k\cos k) \dots \nonumber\\
 &\dots+ (\cos k - \Sm k \sin k) (\sin k + \Sp k \cos k) = 0.
\end{align}
In the following, we review several special cases of the general flow considered above. These are used as reference solutions in the results section.

\subsection{No-slip boundary conditions}
For $S^+ = S^- =0$, the boundary conditions \cref{eq:bcs} reduce to 
$u=0$ at $y=\pm 1$, which are the no-slip boundary conditions that are typically used for liquid wall interactions on the macroscopic scale. This allows to reduce the slip solution  \cref{eq:solNavierSlip} to 
\begin{align}\label{eq:solNoSlip}
 u(t,y) 
 =
 1- y^2 - \sum_{n=1}^\infty A_n \sin (k_n(y+1)) \exp(-k_n^2 t).
\end{align}
In contrast to the problem with Navier slip boundary conditions, the coefficients can be \emph{explicitly} computed giving 
\begin{align} \label{eq:charEqNoSlip}
k_n=n\pi/2, \quad n \in \mathbb{N}.
\end{align}
\change{Hence, the eigenvalues $k_n$ are equally spaced for the case of no-slip boundary condition. This is not true if a finite slip is present at the solid boundary (see below). Hence, a quantitative analysis of the spectrum of hydrodynamic eigenmodes allows to unambiguously distinguish between slip and no-slip \cite{Chen2015}.}\\
\\
Moreover, \eqref{eq:An} reduces to 
\begin{align} \label{eq:AnNoSlip}
 A_n 
 =
\begin{cases} 
4/k_n^3 &\mbox{for $n$ odd}\\
0 & \mbox{for $n$ even.}
\end{cases} 
\end{align}
This solution can, e.g., be found in collections for analytic solutions, such as \cite{Brenn2017}.

\subsection{Channel flow with equal wall slip}
For the case where $\Sp = \Sm =: S$ the stationary solution simplifies to
\begin{align} \label{eq:usEqualSlip}
\overline{u}(y) = 2S +1 - y^2.
\end{align}
In this case, \cref{eq:charEqFactored} gives
\[ (1-k^2 S^2) \sin(2k) + 2kS \cos(2k) = 0. \]
The formula for the series coefficients $A_n$ can be simplified to
\begin{align}\label{eq:AnNavierSlip}
 A_n 
 =
\begin{cases} 
 \frac{4 \sin k_n (\sin k_n + k_n S \cos k_n) }{k_n^3 ( k_n^2 S^2 + S +1)}&\mbox{for $n$ odd}   \\
\hspace{1.8cm} 0 & \mbox{for $n$ even}.
\end{cases} 
\end{align}

\subsection{Free slip at both boundaries}
For $S^\pm \to \infty$ the Navier slip boundary condition is equivalent to the \emph{free slip} boundary condition
\begin{align}
 \partial_y u = 0 \quad \text{at} ~ y=\pm 1.
\end{align}
In this case, we only obtain the unbounded solution
\[  u(t,x) = 2 t,\]
for the overall problem. Even though this solution does not require the computation of any series coefficients, we note for a later comparison that in this case, the solution of the characteristic equation is 
\begin{align}\label{eq:charEqFreeSlip}
 k_n = n \pi, \quad n \in \mathbb{N}.
\end{align}

\subsection{Symmetry and partial slip}\label{sec:symmetry_and_partial_slip}
\change{It is more interesting to consider the case where free-slip holds only on one boundary (either top or bottom). Note that the free-slip condition can also be understood as a symmetry condition. Therefore, the problem is equivalent to a channel of twice the height with equal slip length on both sides.}\\
\\
\change{By rewriting the characteristic equation \eqref{eq:charEq} as}
\[ \frac{\sin 2k}{S^-} + k \frac{S^+}{S^-} \cos 2k = k (S^+ k \sin 2k - \cos 2k), \]
\change{we obtain the form of the characteristic equation in the limit $S^- \to \infty$ and $S^+/S^- \to 0$}
\begin{align*}
0 = k \left(S^+ k \tan 2k -1 \right). 
\end{align*}
\change{Hence, any solution $k>0$ satisfies}
\begin{align}\label{eq:charEq_limiting_case}
S^+ k \tan(2k) = 1. 
\end{align}
\change{We are interested in the asymptotic behavior of the smallest positive solution $k_1$, since it corresponds to the leading eigenmode with the largest decay time. In particular, we consider the limiting cases}
\[ S^+ \to \infty \quad \text{and} \quad S^+ \to 0 \]
\change{in \eqref{eq:charEq_limiting_case}.}
\begin{enumerate}
 \item \change{We assume $S^+ \gg 1$ such that}
 \[ k \tan(2k) = \frac{1}{S^+} \]
 \change{is close to zero. In this case, we may apply a Taylor expansion $k \tan(2k) = 2 k^2 + \mathcal{O}(k^4)$ to find}
 \begin{align}\label{eq:asymptotics_free_slip_large_slip}
 k_1 = \frac{1}{\sqrt{2 S^+}} 
 \end{align}
 \change{asymptotically as $S^+ \to \infty$. In this case, the non-dimensional decay time is proportional to $S^+$}
 \[ \tau_1 = \ln 10/k_1^2 = \frac{\ln 10}{2} \, S^+. \]
 \item \change{Conversely, if $S^+ \to 0$ in \eqref{eq:charEq_limiting_case}, we are looking for solutions of}
 \[ S^+ = \frac{1}{k \tan 2k}=:f(k). \]
 \change{Approximating $f(k)$ linearly near its smallest root $\pi/4$, i.e.}
 \[ f(k) = -\frac{8}{\pi} \left( k - \frac{\pi}{4}\right) + \mathcal{O}((k-\pi/4)^2, \]
 \change{leads to}
 \begin{align}
 k_1 = \frac{\pi}{4}\left( 1 - \frac{S^+}{2} \right)
 \end{align}
 \change{asymptotically as $S^+ \to 0$. So in this case, the (non-dimensional) decay time behaves like}
\begin{align}
\tau_1 = \frac{16 \ln 10}{\pi^2}  \left( 1 - \frac{S^+}{2} \right)^{-2} = \tau_1^0 + \frac{16 \ln 10}{\pi^2} S^+ + \mathcal{O}((S^+)^2),
\end{align}
\change{where}
\[ \tau_1^0 = \frac{16 \ln 10}{\pi^2}  \]
\change{is the non-dimensional decay time for $S^+ \to 0$.}
\end{enumerate}

 \section{An algorithm for the starting flow with slip boundary conditions}
\label{sec:algorithm}
\begin{figure}[H]
\centering
\includegraphics[width=\textwidth]{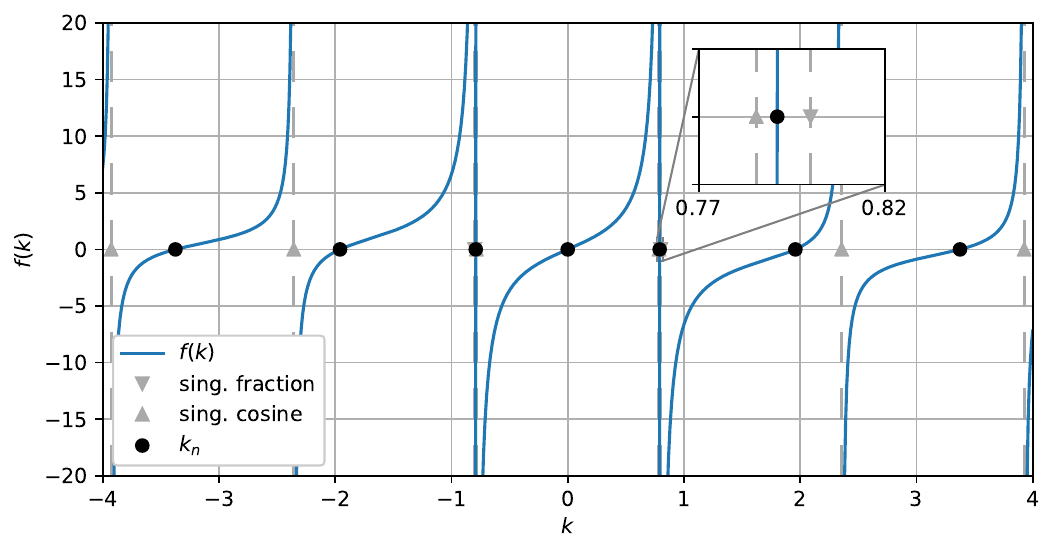}
  \caption{The blue line illustrates the modified characteristic equation \cref{eq:charFunction} for $\Sp = \Sm = 1.25$. The singular points arising from the first and second term of $f$ are marked by upwards and downwards pointing triangles, respectively. The vertical dashed lines mark the location of singularities, while the interval in between two such consecutive points is guaranteed to contain a single zero of $f$. The black dots show the roots of \cref{eq:charFunction}. The inset shows a root that can easily be missed by a manual coefficient computation.}	\label{fig:fPlot}
\end{figure}

In order to evaluate the series solution \cref{eq:utNavierSlip} at some point in time and space, it is necessary to obtain the coefficients $k_n$. We first analyze the properties of \cref{eq:charEq} and based on this analysis, we give an algorithm that obtains all series coefficients in increasing order. We consider \cref{eq:charEq} for $\Sp, \Sm \geq 0$ and distinguish two cases for $\cos(2k)$: 

\paragraph{Case 1:} If $\cos(2k)\neq 0$, we can divide by $\cos(2k)$ and obtain
\begin{align}\label{eq:charEqSpSmStep1}
(1 - \Sp \Sm k^2) \tan(2k) + k (\Sp + \Sm) = 0.
\end{align}
Assuming that $k=\frac{1}{\sqrt{\Sp \Sm}}$ is a root of \cref{eq:charEqSpSmStep1} gives the contradiction 
\begin{align}
 0 < (\Sp + \Sm)/\sqrt{\Sp \Sm} = 0.
\end{align}
Hence, $k=1/\sqrt{\Sp \Sm}$ is not a root of \cref{eq:charEqSpSmStep1}. A similar argument can be made for $k=(2n+1)/4 \pi$.
To proceed, we define the set
\begin{align}
 \Phi = \left\{1/\sqrt{\Sp \Sm} \right\} \cup \left\{ (2n+1)\pi/4 ~ \vert ~ n \in \N_0\right\}
\end{align}
and the function
\begin{align} 
 &f: \R_+ \setminus \Phi \to R \nonumber \\
 &k \mapsto \tan(2k) + \frac{k(\Sp + \Sm)}{1-\Sp \Sm k^2}. \label{eq:charFunction} 
\end{align}
This function is illustrated by the blue line in \cref{fig:fPlot}. The singularities of $f$ are marked by triangles on the $x$-axis, which corresponds to the elements of $\Phi$. The upward and downward-pointing triangles indicate the location of a singularity caused by the tangent-term and the fraction term in \cref{eq:charFunction}, respectively. Hence, the triangles indicate the elements of $\Phi$. The black dots show the locations of the roots of \cref{eq:charFunction}. Between every two points where $f$ is singular, there is a root of $f$. The inset in \cref{fig:fPlot} highlights the case where two singularities are nearly coinciding. Note that for such a case, the corresponding root could easily be missed with a manual approach.
We note that only $k>0$ is admissible for the case of the \change{Sturm-Liouville} problem leading to \cref{eq:utNavierSlip}, see the Appendix of \cite{Matthews2012}. As shown above, the set $\Phi$ does not contain any roots of the characteristic equation \cref{eq:charEq}. Hence, in this case, the roots of \cref{eq:charFunction} are the roots of \cref{eq:charEq}. The elements of $\Phi$ can be ordered such that 
\begin{align}\label{eq:phiOrder}
    0 < \phi_1 < \phi_2 < \dots \qquad \forall \phi_n \in \Phi.
\end{align}
The function $f$ is strictly monotonic in $\R \setminus \Phi$ as
\begin{align}
 f'(k) = \frac{2}{\cos^2(2k)} + (\Sp + \Sm) \frac{1+\Sp \Sm k^2}{(1-\Sp \Sm k^2)^2} > 0 \quad \forall k \in \R_+ \setminus \Phi.
\end{align}
Furthermore, we have for all singularities of $f$:
\begin{equation}
 \lim_{k\downarrow \phi_n} f(k) = -\infty \quad \text{and} \quad \lim_{k\uparrow \phi_n} f(k) = \infty 
  \qquad \forall n \in \N \label{eq:upDownSing}.
\end{equation}
As $\R_+ \setminus \Phi$ is open, it follows from \cref{eq:upDownSing} that there exist $a_n,b_n \in (\phi_n,\phi_{n+1})$ such that $f(a_n)<0$ and $f(b_n)>0$. Hence, the intermediate value theorem applied to the restricted function $\tilde{f}_n:= f|_{[a_n,b_n]}$ guarantees the existence of a root $k_n$ for which $a_n< k_n < b_n$. As $f$ is monotonic so is $\tilde{f}_n$ and consequently the root of $\tilde{f}$ in $(\phi_n, \phi_{n+1})$ is unique. Given the input for the left and right bounds as above the bisection algorithm converges to the root of a strictly monotonic function. A proof of the convergence of the bisection method under the given conditions can be found in \cite[chapter 2, p.51, Theorem 2.1]{Burden2010}.

\paragraph{Case 2:} \change{It follows} from $\cos(2k) = 0$ that $2k=(2n+1)\pi/2$, $n \in \N_0$, as $k>0$. In this case, we also have $\sin(2k) = (-1)^{2n+1}$. Plugging these into \cref{eq:charEq} 
leads to
\begin{align}\label{eq:kExtra}
 k = \frac{2\tilde{n} + 1 }{4} \pi = \frac{1}{\sqrt{\Sp \Sm}}.
\end{align}
Hence, this case gives a single coefficient $k$ if and only if a natural number $\tilde{n} \in \N_0$ exists such that 
\begin{align}\label{eq:kExtraCondition}
 \Sp \Sm = \frac{16}{(2\tilde{n} +1)^2 \pi^2}.
\end{align}
Otherwise, this case does not provide any coefficient at all. 

Considering \cref{fig:fPlot}, case 2 can be illustrated as follows: If the slip length values are continuously changed, the singularity due to the fraction term in \cref{eq:charFunction} is moving on the $x$-axis. When \cref{eq:kExtraCondition} is  satisfied, the singularity due to the fraction term of \cref{eq:charFunction} is coinciding with one of the singularities cause by the tangent-term, i.e. the downward pointing triangle is located at the same location as one of the upward-pointing triangles. Even if \cref{eq:kExtraCondition} is only approximately satisfied, a root of \cref{eq:charFunction} is located between nearly coinciding singular points and could easily be missed if the coefficients $k_n$ would be obtained by a manual approach. Note that this case also occurs for the special case $\Sp = \Sm$.\newline
\newline

To automatically compute all coefficients $k_n$, we propose the following algorithm:
\begin{algorithm}[H]
    \centering
    \caption{Start-up Coefficient Computation (SCC)} 
    \label{alg:startup} 
    \begin{enumerate}
    \item Start: Given $\Sp , \Sm \geq 0$ the algorithm computes the first $N$ coefficients $k_1, \dots k_N$. We start by initializing $\tilde{\Phi} = \left\{ (2n+1)\pi/4 ~ \vert ~ n \in \N_0\right\}$.
    \item The index $\tilde{n}$ of the root caused by the fractional term in \cref{eq:charFunction} is obtained by 
\begin{align}
     m:= \left(\frac{1}{\Sp \Sm} - \left(\frac{1}{\Sp \Sm} \mod \frac{\pi}{4} \right)\right) \big/ \frac{\pi}{4}
    \end{align}
from which we set
\begin{align}
    \tilde{n} := 
        \begin{cases} 
        (m-1)/2 &\mbox{if $m$ is odd}  \\
        (m-2)/2 &\mbox{if $m$ is even.}
        \end{cases} 
    \end{align}
    If $\tilde{n}>N$, then $k_{\tilde{n}}>k_N$ and the root caused by the fractional term does not coincide with any of the first $N$ roots and we continue with step 3. Otherwise, based on the considerations for case 2, \cref{eq:kExtraCondition} is used to identify the case where two singular points of \cref{eq:charFunction} coincide by
\begin{align}\label{eq:mergeCond}
        \vert \pi^2 \Sp \Sm (2\tilde{n} +1)^2  - 16 \vert \leq 3\,\ulp.
    \end{align}
Here \ulp is the unit in the last place.
    If \cref{eq:mergeCond} is not satisfied, we add $1/\sqrt{\Sp \Sm}$ to $\tilde{\Phi}$ and continue with the next step. Otherwise, the root with  index $\tilde{n}$ is set to $1/\sqrt{\Sp \Sm}$ due to \cref{eq:kExtra} and roots with indices $1, \dots, \tilde{n}-1, \tilde{n}+1, \dots N$ are computed in the next step.
    \item All remaining coefficients are obtained using the bisection method to obtain the unique root $k_{i}$ of $f$ in the interval $(\phi_i, \phi_{i+1})$ where $\phi$ are defined in \cref{eq:phiOrder}. 
    The initial values for the bisection method
\begin{align}
        a_i := \phi_i + \ulp \qquad \text{and} \qquad
        b_i := \phi_{i+1} - \ulp
    \end{align}
are used.
\end{enumerate}
\end{algorithm}

Roots of \cref{eq:charFunction} can be located arbitrarily close to singularities, so that in these cases, the precision of the used floating-point arithmetic becomes relevant.
Due to the finite floating point precision available on the hardware, we recommend to increase the used floating point precision when required. For this purpose, we provide two different implementations: one using python that is based on a standard double floating point standard and a C++ that provides an arbitrary precision implementation. The C++ implementation is built using the boost multi-precision library.

 \section{Results}
\label{sec:results}
\subsection{Verification against arbitrarily precise results}

A comparison of our results to coefficient values from literature showed that the provided precision is relevant when comparing coefficients for different values for $\Sp$ and $\Sm$. We refer to \cref{sec:parameterStudies} for a detailed comparison and take this observation as a motivation to shortly compare between different implementations of our algorithm with respect to their accuracy.

To provide a reliable reference, improve computation speed, and extend the range of applicability of the implementation, the algorithm has been implemented using the C++ programming language in combination with the boost arbitrary precision library. With this implementation, it is possible to compute an arbitrary number of coefficients with arbitrary accuracy. The two independent implementations (double-precision python based and C++-based arbitrary precision) will be compared in the following using the arbitrary precision implementation with 50-significant decimal digits. The APA-bisection method is using a $\texttt{rTol} =10^{-40}$, while the double-precision implementation is using the scipy default settings. The precision requirement will become more evident in the comparison to literature values further below.

\begin{figure}[H]
 \centering
 \includegraphics[width=\textwidth]{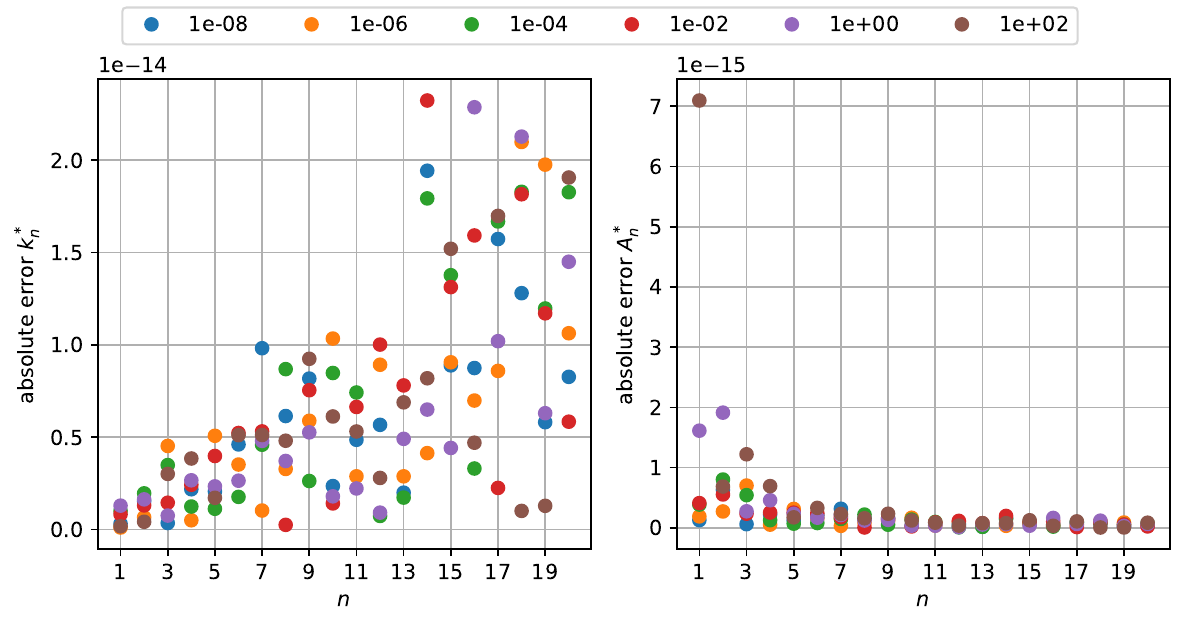}
 \caption{Absolute errors for coefficients from the python implementation for $k^*_n$ and $A^*_n$ for \mbox{$S=10^{-8}, 10^{-6}, 10^{-4}, 10^{-2}, 1, 10^{2}$}.}
 \label{fig:absErrors}
\end{figure}

\begin{figure}[H]
 \centering
 \includegraphics[width=\textwidth]{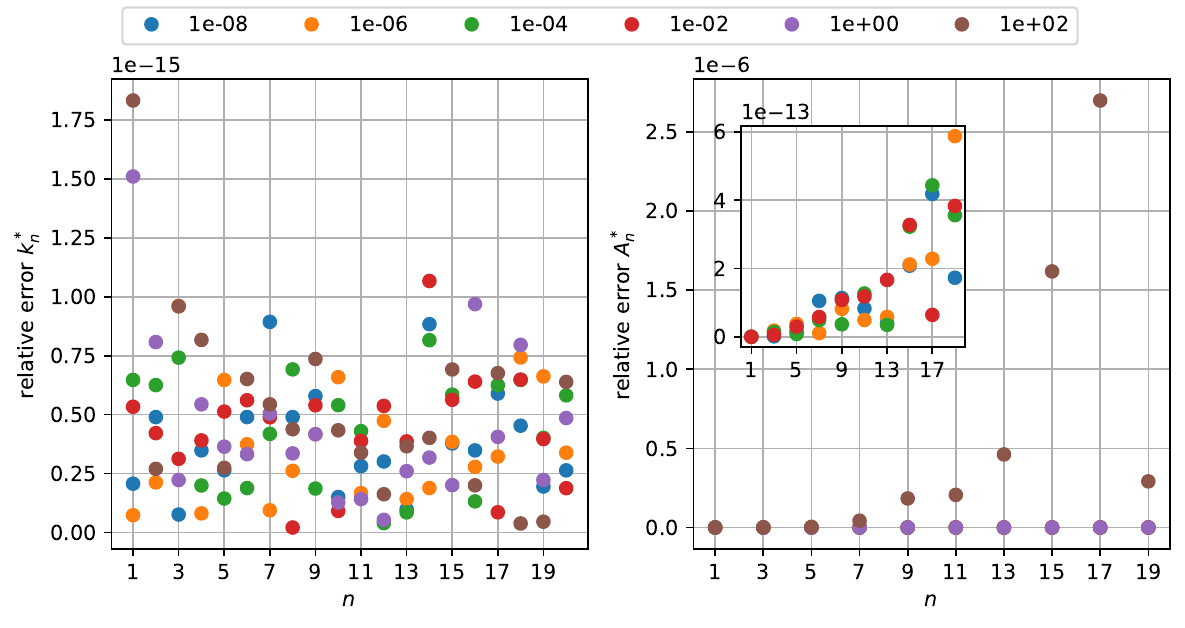}
 \caption{Relative errors of the python implementation for $k^*_n$ and $A^*_n$ for \mbox{$S=10^{-8}, 10^{-6}, 10^{-4}, 10^{-2}, 1, 10^{2}$}.}
 \label{fig:relErrors}
\end{figure}

A comparison of absolute and relative errors for $k^*_n$ and $A^*_n$ for $\Sp = \Sm = S$ can be seen in \cref{fig:absErrors} and \cref{fig:relErrors}, respectively. The errors are obtained using the arbitrary precision implementation in C++ as reference. The comparison includes those coefficients for even $n$ to demonstrate that the accuracy of the non-zero coefficients is the same as for those that are analytically zero.
While the absolute errors in \cref{fig:absErrors}  increase for $k^*_n$ with increasing $n$, the relative error in \cref{fig:relErrors} is between $10^{-15}$ and $10^{-16}$ for all $k^*_n$. For the $A^*_n$-coefficients, the absolute error decreases while the relative error increases with increasing $n$. The fluctuation of the errors can be explained by an accumulation of round-off and cancellation errors as these values are close to the double-precision machine-tolerance. 

The computation of the $A^*_n$ coefficients is based on the computation of $k^*_n$ which show a small though positive absolute error. The error for $A^*_n$ is well below $10^{-15}$ for nearly all coefficients. Furthermore, the trend for the absolute error is inverse to the one for $k^*_n$ as with increasing index the error decreases. This indicates, that the formula for $A^*_n$ is dampening the error introduced by $k^*_n$. 
The relative errors for $k^*_n$ are similar to the absolute errors close to machine tolerance and of the same order as for those coefficients that are analytically zero. Again, all errors are on the same order of magnitude - nearly all of them are between $10^{-15}$ and machine tolerance at $10^{-16}$. These errors are satisfactory for the used double-precision floating point numbers. The seemingly random distribution is likely due to the round-off errors.

The situation is somewhat different for the relative error for $A^*_n$ depicted in \cref{fig:absErrors}. The relative error for the coefficients corresponding to the case for $S=100$ show relative errors on the order of $10^{-6}$ (brown dots). With decreasing $S$ the relative error is on the order of $10^{-13}$, which can be seen in the inset in \cref{fig:relErrors}. This behavior can be explained as follows: for increasing $S$, the coefficients $A^*_n$ quickly decrease in magnitude. Hence, they are barely distinguishable from $0$ and further decrease with increasing index number. This behavior makes arithmetic operations between large $S$ and small $k^*_n$ prone to round-off errors.

\begin{figure}[H]
 \centering
 \includegraphics[width=\textwidth]{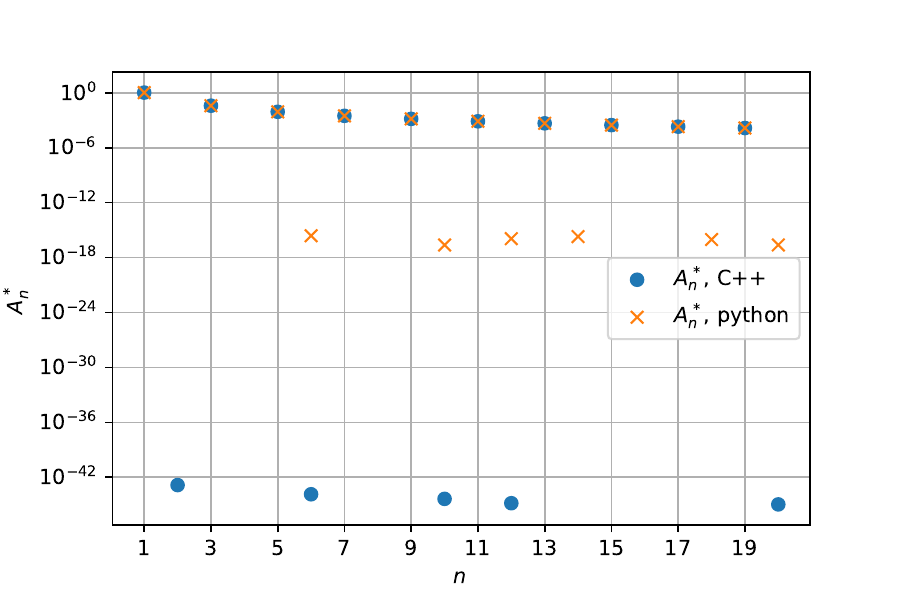}
 \caption{Comparison of $A^*_n$-coefficients between python and C++ implementation for $S=10^{-2}$. Python results can be obtained up to an absolute error of $10^{-16}$, while the C++ implementation provides results with an error smaller than $10^{-40}$. The results for $A^*_n$ reflect this accuracy with absolute values smaller than $10^{-16}$ and $10^{-40}$ for the python and C++ implementations, respectively, whereas it follows analytically that $A^*_{2n}=0 ~\forall n \in \mathbb{N}$, see \cref{eq:AnNavierSlip}.}
  \label{fig:anCompCppPython}
\end{figure}

Based on an accurate computation of $k^*_n$, the $A^*_n$-coefficients are obtained and shown in a semi-logarithmic plot in \cref{fig:anCompCppPython}. It can be seen that the results from the python and C++ implementation show excellent agreement. The numerical results for the coefficients with even index are also shown in the plot. As the python implementation is based on a \texttt{double} precision, the absolute values for the $A^*_n$-coefficients are approximately $10^{-16}$, which corresponds to machine precision for this case. The results using the arbitrary precision C++-implementation have a relative tolerance of $10^{-40}$, which is reflected in results for the coefficients with even index obtained from the C++-implementation. 

\subsection{Verification of Navier slip boundary conditions}
\label{sec:verification}

While the comparison in the former section shows an excellent agreement between the two different implementations of the solution algorithm presented in \cref{sec:algorithm}, this section compares the results of the start-up algorithm to the full direct numerical solution (DNS) of the single phase Navier-Stokes equations between two plates. Here, the fully resolved flow provides an ``external'' reference solution for the implementation. To solve the single phase Navier-Stokes equations in a channel flow geometry, the CFD-framework OpenFOAM extend v.3.1 is used. Specifically, the single phase Navier-Stokes solver \texttt{icoFoam} is applied for this purpose. In the following, the solutions obtained by the DNS of the Navier Stokes equations are referred to as ``CFD-solution'', while the result of the start-up algorithm is referred to as ``analytic solution''.

As \cref{eq:momEq} neglects the influence of the convection term, we consider a verification case with a vanishing Reynolds number. Furthermore, as the solution depends only on time $t$ and the spacial $y$-direction, no changes in the velocity field in $x$-direction have to be expected. Hence, it is sufficient to  only resolve the domain in $y$-direction, where $N_y$ is the number of cells used in the cross section of the channel. As input parameters for the OpenFOAM solver \texttt{icoFoam}, we use a kinematic viscosity of $\nu = \SI{0.1}{\meter \squared \per \second}$ and a channel height of $H=\SI{1}{\meter}$.

To demonstrate the applicability of the start-up algorithm for a wide range of parameters, a large variety of slip lengths ranging from $L=\SI{e-8}{}$ to $\SI{e1}{\meter}$ is considered. To analyze the mesh convergence for each of these slip lengths, mesh resolutions between $N_y=5$ and $N_y=160$ have been used, where the resolution has been increased by a factor of two each time.

\begin{figure}[H]
 \centering
 \includegraphics[width=1.0\textwidth]{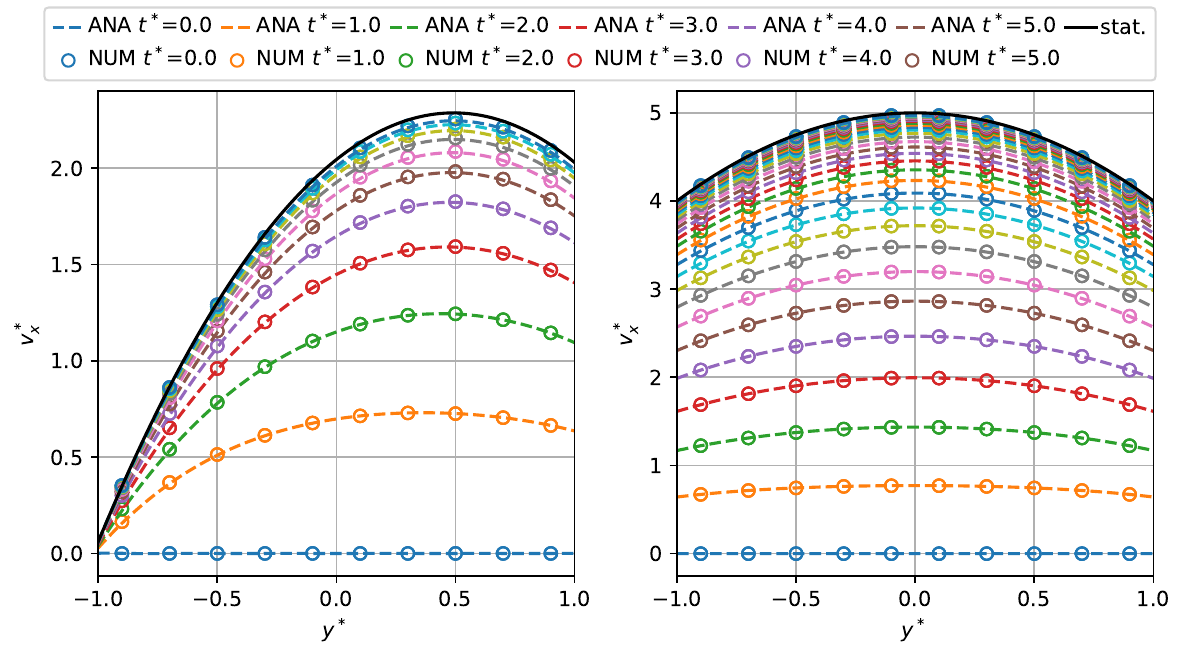}
 \caption{Comparison of non-dimensional results obtained from the analytic solution (ANA) using the start-up algorithm and results from the full numerical solution of the transient Navier-Stokes equations using OpenFOAM (NUM). In the left plot, results for $\Sp=2$ and $\Sm=0.2$ are shown. The right plots illustrates the case for $\Sp=\Sm=2$. The time interval between two subsequent velocity profiles is 1.0. The first six velocity profiles are labeled in the legend.} 
 \label{fig:compVelFields}
\end{figure}

The results of these mesh convergence studies are illustrated in \cref{fig:compVelFields}. The velocity field from the analytic solution (ANA) obtained from the start-up algorithm and results from the full numerical solution of the transient Navier-Stokes equations using OpenFOAM (NUM) are compared. The developing velocity fields are shown for two different length scale ratios. The left figure illustrates the case with $\Sm=0.2$ on the left and $\Sp=2$ on the right wall. The right figure shows results for the case with an equal slip length of $\Sp = \Sm =: S = 2$.  All quantities have been scaled using \cref{eq:scales}. The results obtained from the full numerical solution of the velocity field are depicted by empty circles. Note that only a subset of the CFD-solution has been plotted to allow for a better comparison to our solution, which is shown as dashed lines. The different times at which the velocity field is shown is indicated by different colors. The colors repeat periodically with increasing time. The different evaluation times vary between the initial time at $t^*=0$ to nearly twice the characteristic time of the no slip solution ($R^2/\nu$), which is $2.5$ in this case.

Starting from an initially identically zero velocity field, the velocity increases in magnitude over time and converges to the stationary solution (continuous black line).  Comparing the two start-up solutions on the left and on the right one can observe the influence of the different slip lengths on the solution. With increasing slip length, the slip at the channel walls at $y=\pm1$ increases, as well as the maximum velocity in the center of the channel. Both effects contribute to an overall increased volume flux through the channel. Also note that the characteristic time scale increases with increasing $S$. This effect is further detailed below. Overall, an excellent agreement between numerical and analytical velocity field can be observed for all times including the limit for the stationary solution.

\begin{figure}[H]
 \centering
 \includegraphics[width=\textwidth]{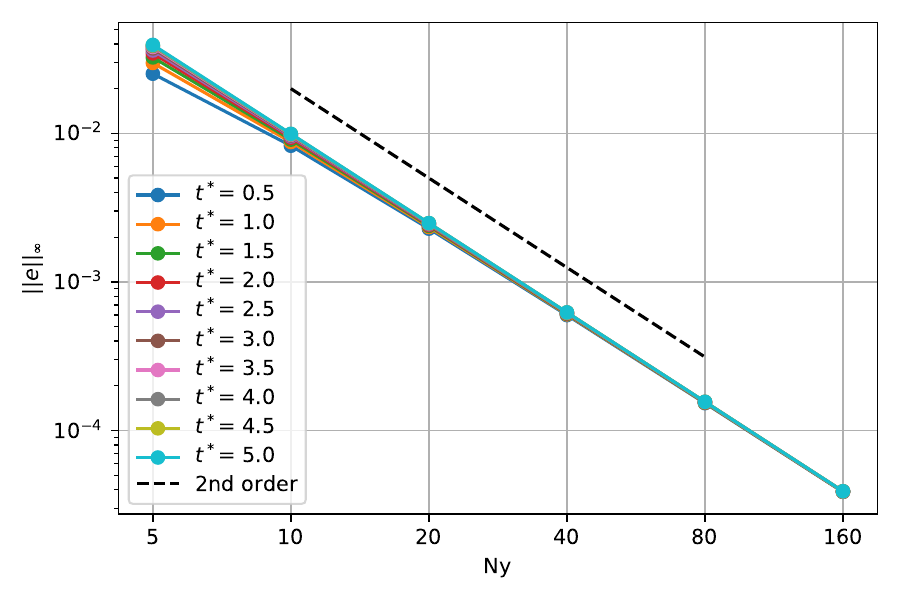}
 \caption{Mesh convergence study using OpenFOAM's \texttt{icoFoam} solver with Navier slip boundary conditions for $\Sp=2\times10^{-4}$ and $\Sm=2\times 10^{-2}$. The results of the start-up algorithm have been used as analytic reference solution.}
 \label{fig:meshConv} 
\end{figure}

A mesh convergence study has been performed for the parameter set described above. An exemplary result for  $\Sp=2\times10^{-4}$ and $\Sm=2\times 10^{-2}$ is shown in \cref{fig:meshConv}. The horizontal axis shows the mesh resolution in the cross section of the channel. The vertical axis depicts the error of the CFD-solution that is defined by
\begin{align}\label{eq:errorNorm}
 e = \norm{u_{\text{num}} - u_{\text{ana}}}_\infty
\end{align}
where $u_{\text{num}}$ is the vector containing the velocities in $x^*$-direction obtained from the CFD-solution and $u_{\text{ana}}$ is the vector containing the results of the analytic solution. 

The plot contains results for a variety of dimensionless times between $t=\SI{0.5}{}$ and $\SI{4.5}{}$. Each of the continuous lines corresponds to the mesh convergence for different meshes at a certain dimensionless time. Each dot on the line corresponds to a certain mesh resolution $N_y$ where the error has been computed by applying \cref{eq:errorNorm}. A comparison of slope of the error curves to the 
second order (dashed) reference curve shows a second order convergence of the CFD-results. This is also the expected convergence order based on the choice of the spacial discretization schemes of the CFD-solution. 
Similar results have been obtained for arbitrary combinations of $\Sp$ and $\Sm$ with $\Sp, \Sm \in \{0, 2 \cdot 10^{-8}, 2 \cdot 10^{-4}, 0.02, 2, 20\}$. 
Hence, it can be concluded that not only the two implementations of the start-up algorithm are consistent with each other, but are also consistent with the numerical solution of the full continuum mechanical problem. It should be noted that with increasing $\Sp$ and $\Sm$ the problem becomes more challenging for a CFD-solution as required accuracy for the solutions of the involved linear solvers increases. 
\subsection{Parameter studies for $\Sp$ and $\Sm$}
\label{sec:parameterStudies}

\begin{figure}[ht]
\centering
 \includegraphics[width=\textwidth]{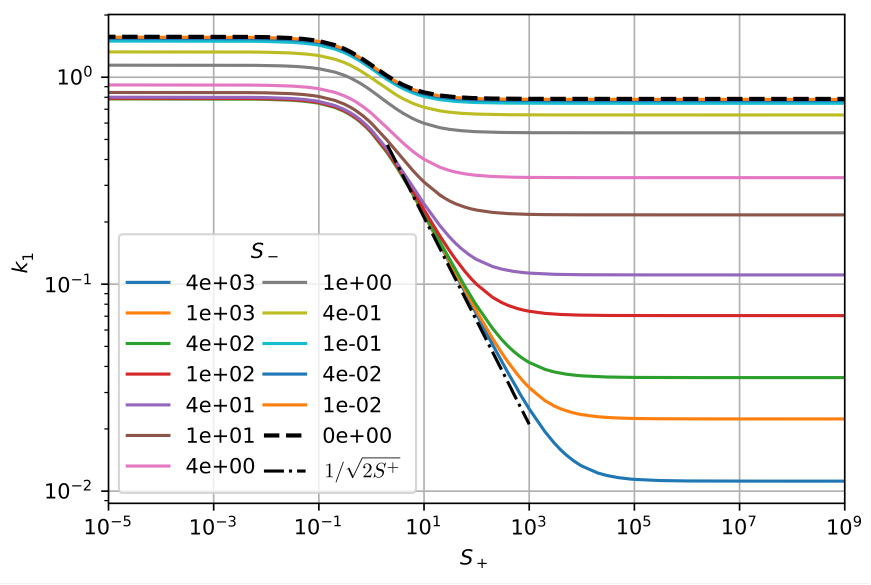}
 \caption{Comparison of the solutions for the first root of \eqref{eq:charEq} by application of \cref{alg:startup}. The horizontal axis depicts the values for $\Sp$, and the different colors of the curves represent the different values for $\Sm$.}
 \label{fig:k1_coeff} 
\end{figure}    
\change{As discussed in \cref{sec:analyticSolution},  the first root $k_1$ of the characteristic equation \cref{eq:charEq} (i.e.\ the square-root of the leading eigenvalue) is central for the characterstic time-scale of the start-up flow. Hence,} the dependence of $k_1$ on $\Sp$ and $\Sm$ is a valuable information. This dependence is illustrated in \cref{fig:k1_coeff}. Without loss of generality (we could switch the placement of $\Sp$ and $\Sm$ in \cref{eq:charEq}), we place $\Sp$ on the x-axis and use $\Sm$ as parameter for different curves. The horizontal axis depicts the variation of $\Sp$ for approximately ten orders of magnitude. The different colored lines show the value of $k_1$ for different values of $\Sm$. The values for $\Sm$ vary between $\SI{e2}{}$ and $\SI{e-2}{}$ and have been selected to give a good representation of the $k_1$ dependence.

Several limit cases can be seen. Firstly, fixing the value for $\Sm$ the value for $k_1$ levels off at a constant value when increasing or decreasing the value of $\Sp$. 
Secondly, if $\Sm$ is approaching zero, the different curves approach the curve for $\Sm=0$. This limit curve is illustrated by a dashed line in \cref{fig:k1_coeff} and represents the maximum for values of $k_1$. 

All curves show a transition region between their two limit values that begins at $\Sp\approx 10^{-1}$. This transition region increases in width with increasing $\Sm$. In this region and for much greater values of $\Sm$ than shown in \cref{fig:k1_coeff} we find an excellent agreement with the asymptotic power law \eqref{eq:asymptotics_free_slip_large_slip} derived in Section~\ref{sec:analyticSolution} in the limit $S^- \to \infty$ and $S^+ \gg 1$, i.e.\
\begin{align*}
    k_1 = \frac{1}{\sqrt{2 \Sp}}.
\end{align*}
This relation is shown by the dash-dot line in \cref{fig:k1_coeff} and the transition region is well described by this relation for all curves with $\Sm > 10$.

\begin{figure}[H]
\centering
 \includegraphics[width=\textwidth]{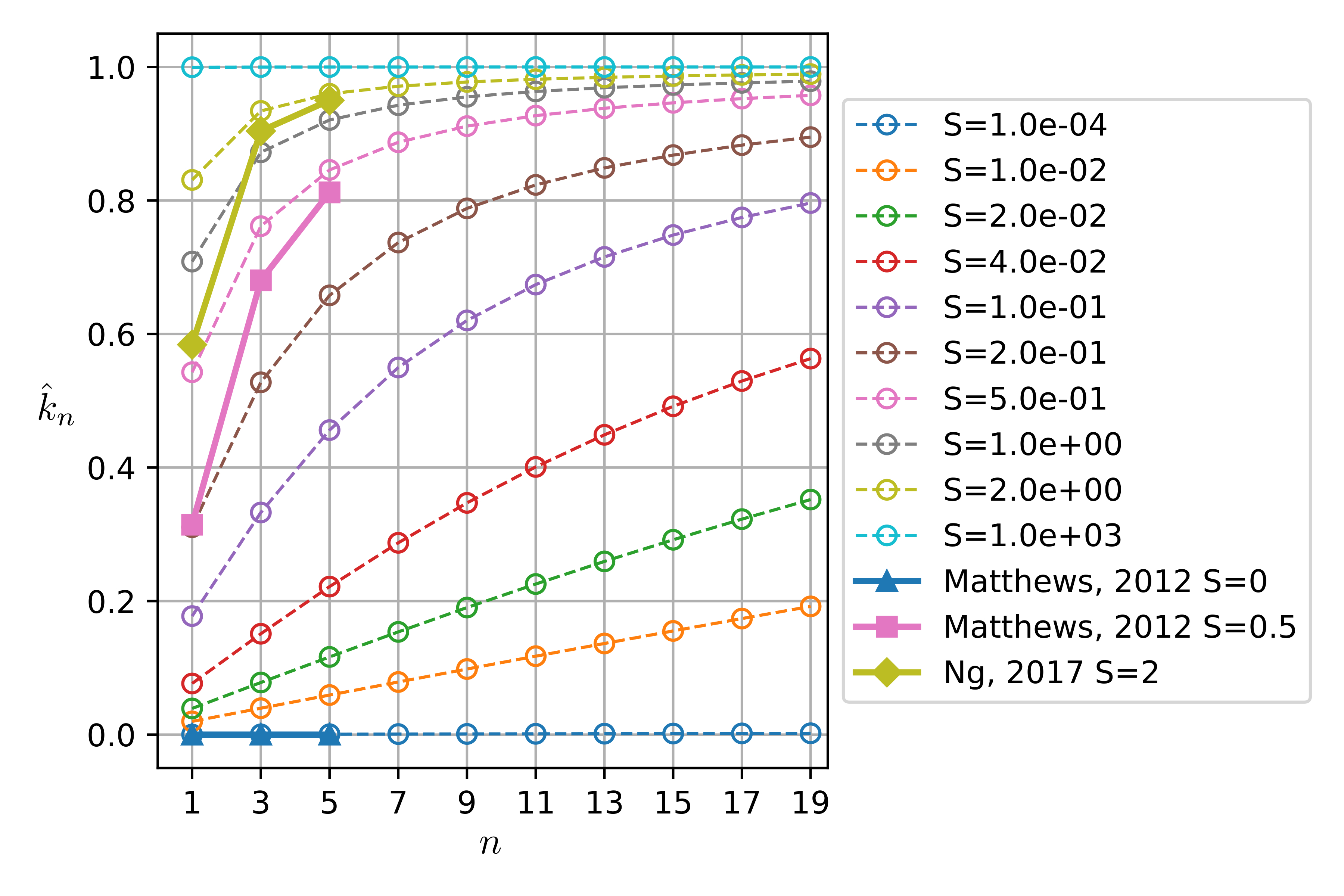}
 \caption{Comparison of the coefficients \change{$\hat{k}_n$} for various values of $S$ for the case $\Sp=\Sm=S$. The limit cases for the no slip and free slip cases are given by \cref{eq:charEqNoSlip} and \cref{eq:charEqFreeSlip}, respectively. The lines have been added to guide the eye.}
 \label{fig:sKnComp_scaled} 
\end{figure}

\paragraph{Comparison with data from the literature:} To improve the comparison, \change{we consider the rescaled coefficients}
\begin{align}
 \change{\hat{k}_n =  (n\pi/2- k_n)/(\pi/2)= n - 2 k_n/\pi.}
\end{align}
This scaling yields the normalized difference to the coefficients for the no slip solution \cref{eq:charEqNoSlip}. Applying this scaling yields the coefficients as depicted in \cref{fig:sKnComp_scaled}. The coefficients vary between 0 and 1 as can be expected for a normalized quantity. The line connecting several coefficients does not indicate intermediate values but has been added to group the coefficients for certain values of $S$. For $S\to 0$, the coefficients \change{$\hat{k}_n$} approach $0$, which corresponds to the limit case with no slip \cref{eq:charEqNoSlip}. For a diverging $S$, the coefficients approach unity, which corresponds to the limit case for free slip boundary conditions and therefore \cref{eq:charEqFreeSlip}. It can be seen that the coefficients and thereby the results are converging to the two limit cases. Furthermore, differences in the solutions are hard to quantify for extremely large or small values of $S$.

\cref{fig:sKnComp_scaled} also shows a comparison to coefficients from literature. In \cite{Matthews2012} five coefficients are given for $S=0$ and $S=0.5$, and in \cite{Ng2017} five coefficients are provided for the case $S=2$. However, as $A_n=0$ if $n$ is even, two of those coefficients are not required for the computation of the velocity field and hence have been omitted in \cref{fig:sKnComp_scaled}. The coefficients from literature are shown in the same color as the corresponding results (for the same slip value) obtained by the C++-implementation of the start-up algorithm. While the results coincide for the analytically available case $S=0$, a significant difference for the two other cases $S=0.5, 2$ can be observed. Even though the results provided in \cite{Matthews2012,Ng2017} agree with the results obtained here up to the last significant digit, the precision provided in \cite{Matthews2012,Ng2017} is not sufficient to accurately distinguish between different slip lengths. In this case, the first coefficient for $S=0.5$ from \cite{Matthews2012} coincides with the results for $S=0.2$, with a similar error for the results with $S=2$ from \cite{Ng2017}. Apparently, with increasing $n$, the error decreases. \change{For future reference, a selection of coefficients for a wide range of values for $S$ is given in \cref{app:coefficients} with 16 significant digits in a normalized floating point format.}

\subsection{Characteristic time-scales}
\begin{figure}[H]
 \centering
 \includegraphics[width=0.8\textwidth]{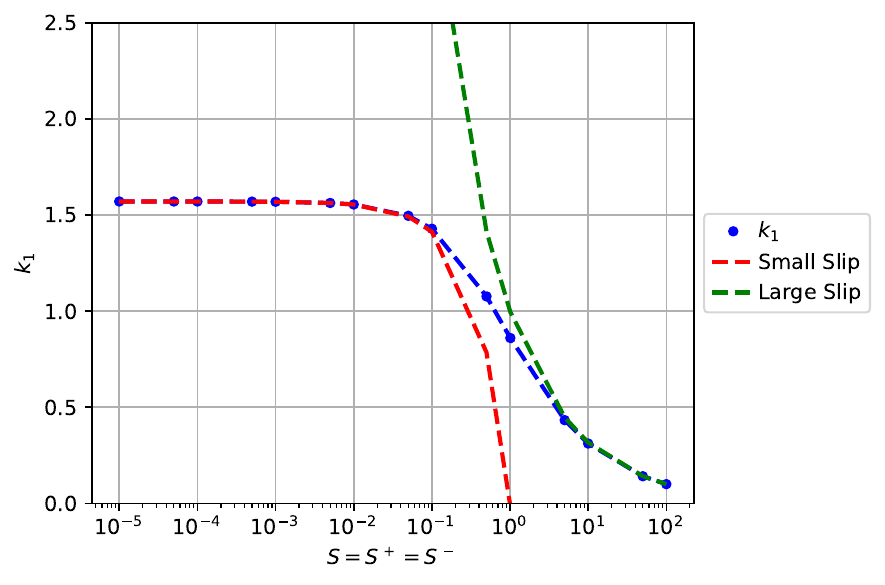}
 \caption{Leading coefficient $k_1=\sqrt{\lambda_1}$ as a function of $S=\Sp=\Sm$. Comparison with asymptotic formulas for small and large slip.}
 \label{fig:k1_vs_S}  
\end{figure} 
As reported in \cite{Kaoullas2015, Avramenko2015, Ng2017, AbouDina2020}, the time scale required to reach a certain fraction of the stationary velocity field increases with increasing slip length \change{($S=S^+=S^-$)} on the channel walls. However, these time scales have not been quantified like in the case of no-slip \cite{Patience1989}.\\
\\
\change{We compare the time scale obtained from the Fourier series solution with the decay time (for $\alpha=0.1$) of the leading eigenmode}
\begin{align}
\tau_1 = \frac{\ln 10}{k_1^2}. 
\end{align}
\change{This can be expected to  be a very good approximation to the start-up time. In particular, from the considerations in Section~\ref{sec:symmetry_and_partial_slip}, we know asymptotic expansions of the decay time in the limit of very small and very large slip. The results for the setup with partial slip and symmetry conditons can be transformed to the case of equal partial slip.\footnote{This transformation  basically involves scale all quantities with $2R$ instead of $R$, i.e.\ by making use of the mirror symmetry.} By this procedure, we obtain}
\begin{align}\label{eqn:startup_time_small_slip_formula}
k_1 = \frac{\pi}{2} \left(1 - S \right) \quad \Rightarrow \quad \tau_1 = \frac{4 \ln 10}{\pi^2} \left(1 - S \right)^{-2} = \frac{4 \ln 10}{\pi^2} \left(1 + 2S \right) + \mathcal{O}(S^2).
\end{align}
\change{in the limit of small slip ($S \to 0$) and}
\begin{align}\label{eqn:startup_time_large_slip_formula}
k_1 = \frac{1}{\sqrt{S}} \quad \Rightarrow \quad \tau_1 = \ln(10) \, S
\end{align}
\change{in the limit of large slip ($S \to \infty$).}

\begin{figure}[H]
 \centering
 \includegraphics[width=0.8\textwidth]{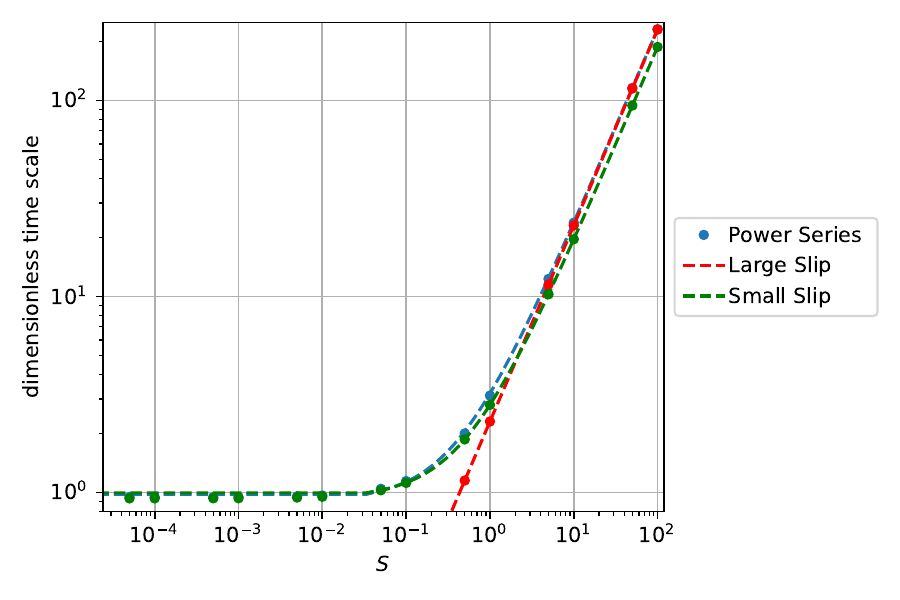}
 \caption{Characteristic relaxation time (to reach 90\% of the stationary velocity) for startup flow with slip. \change{Comparison with asymptotic formulas for the small and large slip limits.}}
 \label{fig:charTimes}  
\end{figure}

\paragraph{Numerical results:} Using our algorithm, a parameter study for different values of \change{$S=S^+=S^-$} has been performed yielding \change{the coefficient $k_1$ and the} dimensionless time to reach $90\%$ of the stationary velocity.\\
A comparison of the (up to numerical precision) exact value of $k_1$ with the asymptotic formulas is shown in \cref{fig:k1_vs_S}. It is found that $k_1$ is well-described by the asymptotic formulas for $S \gtrsim 3$ and $S \lesssim 0.1$. Moreover, a regime transition from large to small slip is visible in between where none of the asymptotic formulas is applicable.\\
The results for the dimensionless time-scales can be seen in \cref{fig:charTimes}.  Here, we have compared the maximum velocity in the center of the channel.
\change{Again, it is found that the numerical results for the start-up time agree very well with the asymptotic formulas for $S \lesssim 0.1$ and $S \gtrsim 3$. In between these values, we observe a non-linear relation between slip and start-up time and the asymptotic formulas show some deviations. However, the overall agreement with the small slip asymptotics \eqref{eqn:startup_time_small_slip_formula} \emph{extended} to arbitrary $S >0$ (neglecting higher-order terms) over the entire parameter range is remarkable.}

 \section{Conclusion}
Navier slip boundary conditions are applied to a wide variety of different flow problems and analytic solutions for corresponding cases are relevant, among others, as reference solutions for CFD-solvers. 
An algorithm has been presented that provides an automated and robust solution to compute the required \change{eigenmodes and expansion coefficients} for the general case of different slip lengths on the channel walls from the corresponding non-linear characteristic equation. Two implementations, one limited to double-precision and one that can provide arbitrarily accurate results are presented. A comparison confirms that both implementations deliver the same results up to standard double precision machine tolerance. Furthermore they are consistent with the full direct numerical solution of the underlying hydrodynamic problem. The arbitrary precision implementation can provide benchmark results with arbitrary accuracy even for cases with extremely small or large ratios between the slip length and the channel height. The algorithm can be conveniently used as a black box, meaning that no additional work of the user is required for the computation of the coefficients. This replaces the tedious and error prone manual solution, especially in the case of nearly coinciding singularities. \change{Moreover, the algorithm was used to verify the derived asymptotic formulas for the leading eigenmode and the characteristic time-scale in the limit of small and large slip.}

\section*{Acknowledgements}
Funded by the Deutsche Forschungsgemeinschaft (DFG, German Research Foundation) – Project-B02 265191195 – SFB 1194.
Calculations for this research were conducted on the Lichtenberg high performance computer of the TU Darmstadt.

\appendix
\section{Reference coefficients}   
\label{app:coefficients}

The following tables provide highly accurate results for the coefficients $k^*_n$ and $A^*_n$ computed with the arbitrary precision arithmetic implementation introduced above and are intended to extend the coefficients available in \cite{Matthews2012, Ng2017} to a much wider range of parameters. For each table, the row gives the corresponding value of $S$ between $10^{-9}$ and $10^2$, while the column gives the value of the coefficient with indices between $1$ and $19$. The values are provided for the first 17 digits with rounding to nearest. These results are expected to be sufficiently accurate for a computation with standard double-precision. Coefficients for which analytically $A^*_n=0$ have been omitted. Note that the coefficients can also be used for the computation of a starting Couette flow as noted in \cite{Ng2017}. 

\newpage
\pdfpagewidth=9in \pdfpageheight=11.69in

\begin{table}[H] 
  \caption{Coefficients $k^*_n$ for different length ratios $S$, part 1.}
\hspace{-4.0cm}
  \begin{tabular}{llllll}
\toprule
{} S &                     n=1 &                   n=3 &                    n=5 &                     n=7 &                    n=9 \\
\midrule
1e-09 &  1.5707963252241004e00 & 4.7123889756723010e00 & 7.8539816261205013e00 & 1.0995574276568702e01 & 1.4137166927016903e01 \\
1e-08 &  1.5707963110869334e00 & 4.7123889332608009e00 & 7.8539815554346672e00 & 1.0995574177608535e01 & 1.4137166799782401e01 \\
1e-07 &  1.5707961697152797e00 & 4.7123885091458391e00 & 7.8539808485763984e00 & 1.0995573188006958e01 & 1.4137165527437517e01 \\
1e-06 &  1.5707947560001405e00 & 4.7123842680004220e00 & 7.8539737800007030e00 & 1.0995563292000984e01 & 1.4137152804001266e01 \\
1e-05 &  1.5707806189887081e00 & 4.7123418569661553e00 & 7.8539030949436954e00 & 1.0995464332921390e01 & 1.4137025570899302e01 \\
1e-04 &  1.5706392628699011e00 & 4.7119177886406973e00 & 7.8531963145044754e00 & 1.0994474840523223e01 & 1.4135753366758927e01 \\
1e-03 &  1.5692271009819729e00 & 4.7076813338280239e00 & 7.8461356593167482e00 & 1.0984590139197781e01 & 1.4123044835202544e01 \\
1e-02 &  1.5552451292561666e00 & 4.6657651417272481e00 & 7.7763740778469526e00 & 1.0887130102147713e01 & 1.3998089735155082e01 \\
1e-01 &  1.4288700112140771e00 & 4.3058014131192230e00 & 7.2281097716272491e00 & 1.0200262588295905e01 & 1.3214185683842919e01 \\
1e00  & 8.6033358901937973e-01 & 3.4256184594817283e00 & 6.4372981791719468e00 & 9.5293344053619631e00 & 1.2645287223856643e01 \\
1e01  & 3.1105284820029772e-01 & 3.1730971766928695e00 & 6.2990593598956464e00 & 9.4353759757608469e00 & 1.2574323161037869e01 \\
1e02  & 9.9833638551126355e-02 & 3.1447725231101660e00 & 6.2847764523279794e00 & 9.4258388739020980e00 & 1.2567166338520057e01 \\
\bottomrule
\end{tabular}
 \end{table} 

\begin{table}[H] 
  \caption{Coefficients $k^*_n$ for different length ratios $S$, part 2.}   
\hspace{-4.0cm}
  \begin{tabular}{llllll}
\toprule
{}S &                   n=11 &                   n=13 &                   n=15 &                   n=17 &                  n=19 \\
\midrule
1e-09 & 1.7278759577465102e01 & 2.0420352227913305e01 & 2.3561944878361505e01 & 2.6703537528809704e01 & 2.9845130179257907e01 \\
1e-08 & 1.7278759421956270e01 & 2.0420352044130137e01 & 2.3561944666304001e01 & 2.6703537288477868e01 & 2.9845129910651735e01 \\
1e-07 & 1.7278757866868077e01 & 2.0420350206298636e01 & 2.3561942545729195e01 & 2.6703534885159755e01 & 2.9845127224590314e01 \\
1e-06 & 1.7278742316001548e01 & 2.0420331828001832e01 & 2.3561921340002115e01 & 2.6703510852002395e01 & 2.9845100364002679e01 \\
1e-05 & 1.7278586808877492e01 & 2.0420148046856024e01 & 2.3561709284834961e01 & 2.6703270522814361e01 & 2.9844831760794289e01 \\
1e-04 & 1.7277031893273573e01 & 2.0418310420129149e01 & 2.3559588947387645e01 & 2.6700867475111043e01 & 2.9842146003361329e01 \\
1e-03 & 1.7261499809036966e01 & 2.0399955122374212e01 & 2.3538410836847422e01 & 2.6676867014042465e01 & 2.9815323715490695e01 \\
1e-02 & 1.7109307259726943e01 & 2.0220834187410407e01 & 2.3332718796715039e01 & 2.6445005751843368e01 & 2.9557735806876639e01 \\
1e-01 & 1.6259361225504154e01 & 1.9327034291602711e01 & 2.2410848328443429e01 & 2.5506382988977435e01 & 2.8610581936555896e01 \\
1e00  & 1.5771284874815882e01 & 1.8902409956860023e01 & 2.2036496727938566e01 & 2.5172446326646664e01 & 2.8309642854452012e01 \\
1e01  & 1.5714326801776345e01 & 1.8854859544307704e01 & 2.1995694888011272e01 & 2.5136719451598626e01 & 2.8277870201784616e01 \\
1e02  & 1.5708599861836207e01 & 1.8850086423035542e01 & 2.1991603294103655e01 & 2.5133139109756151e01 & 2.8274687555520739e01 \\
\bottomrule
\end{tabular}
 \end{table} 
\newpage
\begin{table}[H] 
  \caption{Coefficients $A^*_n$ for different length ratios $S$, part 1.}
\hspace{-4.3cm}
  \begin{tabular}{llllll}
\toprule
{}S &                   n=1 &                   n=3 &                   n=5 &                   n=7 &                   n=9 \\
\midrule
1e-09 & 1.0320491039264819e00 & 3.8224040886165996e-02 & 8.2563928314118538e-03 & 3.0088895158206464e-03 & 1.4157052180061476e-03 \\
1e-08 & 1.0320491225033654e00 & 3.8224041574198647e-02 & 8.2563929800268752e-03 & 3.0088895699806220e-03 & 1.4157052434888137e-03 \\
1e-07 & 1.0320493082721889e00 & 3.8224048454517968e-02 & 8.2563944661726216e-03 & 3.0088901115771622e-03 & 1.4157054983129653e-03 \\
1e-06 & 1.0320511659590730e00 & 3.8224117256988932e-02 & 8.2564093271836596e-03 & 3.0088955272211137e-03 & 1.4157080463036414e-03 \\
1e-05 & 1.0320697426929779e00 & 3.8224805209473876e-02 & 8.2565578926514244e-03 & 3.0089496515156807e-03 & 1.4157335011263999e-03 \\
1e-04 & 1.0322554965384569e00 & 3.8231677511858519e-02 & 8.2580390830709390e-03 & 3.0094876799708495e-03 & 1.4159855409592672e-03 \\
1e-03 & 1.0341116856423846e00 & 3.8299678307806458e-02 & 8.2724045924216522e-03 & 3.0145465588273398e-03 & 1.4182551555536466e-03 \\
1e-02 & 1.0525387018317054e00 & 3.8907643838938369e-02 & 8.3717196531905725e-03 & 3.0334116162406537e-03 & 1.4164048125480716e-03 \\
1e-01 & 1.2237769119142639e00 & 3.8981626729527902e-02 & 6.5284927034244883e-03 & 1.7608409260036317e-03 & 6.0908996408140775e-04 \\
1e00  & 2.2923516074712986e00 & 7.2446719680633780e-03 & 3.4520033975000920e-04 & 4.9806482610557783e-05 & 1.2218509338424761e-05 \\
1e01  & 6.4284132507787826e00 & 1.2300539013759826e-04 & 4.0223434904942518e-06 & 5.3422924776366766e-07 & 1.2715541771687607e-07 \\
1e02  & 2.0033283427296421e01 & 1.2991835173483862e-06 & 4.0784935589731436e-08 & 5.3753972617863454e-09 & 1.2759845080602730e-09 \\
\bottomrule
\end{tabular}
 \end{table} 

\begin{table}[H] 
  \caption{Coefficients $A^*_n$ for different length ratios $S$, part 2.}
\hspace{-4.3cm}
  \begin{tabular}{llllll}
\toprule 
\centering
{}S &                    n=11 &                   n=13 &                     n=15 &                    n=17 &                     n=19 \\
\midrule
1e-09 & 7.7539376703717627e-04 & 4.6975380242443397e-04 & 3.0579232708932778e-04 & 2.1006495093150440e-04 & 1.5046640966999284e-04 \\
1e-08 & 7.7539378099424116e-04 & 4.6975381087998306e-04 & 3.0579233259357290e-04 & 2.1006495471265872e-04 & 1.5046641237837497e-04 \\
1e-07 & 7.7539392056283482e-04 & 4.6975389543373275e-04 & 3.0579238763451399e-04 & 2.1006499252286892e-04 & 1.5046643946100329e-04 \\
1e-06 & 7.7539531604319716e-04 & 4.6975474079711627e-04 & 3.0579293789293681e-04 & 2.1006537049169286e-04 & 1.5046671016800416e-04 \\
1e-05 & 7.7540925028942885e-04 & 4.6976317701963175e-04 & 3.0579842537832575e-04 & 2.1006913685210369e-04 & 1.5046940530977386e-04 \\
1e-04 & 7.7554653701934284e-04 & 4.6984579812085834e-04 & 3.0585179035765763e-04 & 2.1010546768392832e-04 & 1.5049516391538000e-04 \\
1e-03 & 7.7671389919512998e-04 & 4.7049797723244872e-04 & 3.0623454555781676e-04 & 2.1033560400141092e-04 & 1.5063358634863930e-04 \\
1e-02 & 7.6848100398766675e-04 & 4.6036920630577967e-04 & 2.9582993679644034e-04 & 2.0027735474255085e-04 & 1.4115459236280014e-04 \\
1e-01 & 2.4857199236583632e-04 & 1.1458746030383859e-04 & 5.8044230570411602e-05 & 3.1693557376343359e-05 & 1.8393628921613791e-05 \\
1e00  & 4.0667411118106691e-06 & 1.6483518052363482e-06 & 7.6658838066193622e-07 & 3.9451567742351955e-07 & 2.1943484795376281e-07 \\
1e01  & 4.1724166865955799e-08 & 1.6780663383512985e-08 & 7.7673491366614353e-09 & 3.9851199782459977e-09 & 2.2119065233304189e-09 \\
1e02  & 4.1817188411938867e-10 & 1.6806640809123729e-10 & 7.7761827471564228e-11 & 3.9885897807771079e-11 & 2.2134281658204133e-11 \\
\bottomrule
\end{tabular}
 \end{table} 

\newpage
\pdfpagewidth=8.27in \pdfpageheight=11.69in

\eject

\end{document}